\documentstyle[12pt]{article}
\begin{document}
\textwidth 19cm
\textheight 23cm
\begin{flushright}
FTUAM 95/33
\end{flushright}
\vspace{2cm}
\begin{center}
{\LARGE  Thermodynamic Features of Black Holes \\ Dressed with
Quantum Fields}
\end{center}
\begin{center}
David Hochberg\footnote{Based on a talk given at the Third Workshop
on Quantum Field Theory under the Influence of External Conditions,
Leipzig, Germany, 18-22 September, 1995.}
\end{center}
\begin{center}
{\sl Departamento de F\'isica Te\'orica, C-XI,
Universidad Aut\'onoma de Madrid\\
Cantoblanco, 28049 Madrid, Spain}
\end{center}
\begin{abstract}
The thermal properties of black holes in the presence of
quantum fields can be revealed through solutions of the
semi-classical Einstein equation. We present a brief but
self-contained review of the main features of the semi-classical
back reaction problem for a black hole in the microcanonical
ensemble. The solutions, obtained for conformal scalars, massless
spinors and U(1) gauge bosons, are used to calculate the
$O(\hbar)$ corrections to the temperature and thermodynamical
entropy of a Schwarzschild black hole. In each
spin case considered, the entropy
corrections $\Delta S(r)$, are positive definite and monotone
increasing with increasing distance $r$ from the hole, and are of the same
order as the naive flat space radiation entropy.
\end{abstract}
\vfill\eject
\section{Introduction}
The physics of black holes provides a fertile ground in which the
confluence of gravitation, quantum mechanics and thermodynamics
takes place. Progress in our understanding of the thermal features
of black holes demands a deeper understanding of the relationships
among the state functions of black holes in thermodynamic equilibrium
with quantized matter.
A black hole can exist in thermodynamical equilibrium provided it is
surrounded by radiation with a suitable distribution of stress-energy.
In effect, one studies the consequences of coupling the black hole
with its own Hawking radiation. In the semi-classical approach, such
radiation is characterized by the expectation value of a stress energy
tensor obtained by renormalization of a quantum field on the classical
spacetime geometry of a black hole. Using such a stress tensor as a
source in the semi-classical Einstein equation defines the back
reaction problem.

In this talk, we shall review briefly the steps involved in the
1-loop back reaction problem for black holes placing
special emphasis on the need to impose boundary conditions and the
issue of the perturbative validity of the modified black hole metric.
When then go on
to use the solutions so
obtained to calculate the corrections to the temperature at
spatial infinity, $T_{\infty}$, and to the entropy $\Delta S$ by
which quantum fields of spin $0,1/2,1$ augment the usual
Bekenstein-Hawking entropy. One finds that $T_{\infty} \neq T_H =
\frac{\hbar}{8\pi M}$, while $\Delta S \geq 0$ has a global minimum
at the renormalized event horizon and is
monotonically increasing for increasing
distance from the black hole.
The topics sketched here can be found in more detail in
\cite{York85, HK93, HKYa, HKYb}.
\section{Mechanics of the 1-loop Back-reaction}
We begin by choosing the background spacetime.
We therefore suppose we have solved the classical vacuum Einstein equation
\begin{equation}
{\hat G}_{\mu \nu}({\hat g}) = 0,
\end{equation}
for ${\hat g}_{\mu \nu}$, the classical metric. We now inject, or
superimpose, a collection of quantum fields on this classical
background and solve the semi-classical field equation
\begin{equation}
G_{\mu \nu} ({\hat g} + \delta g) = 8\pi \,
<T_{\mu \nu} (\phi, \psi, A_{\mu}, \cdots ) >_{\hat g},
\end{equation}
where $<T_{\mu \nu}>_{\hat g}$ represents the 1-loop quantum
stress-energy tensor for the fields $\phi, \psi, A_{\mu}, ...$
(scalars, spinors, vector bosons, etc.)
renormalized over the background spacetime whose metric is ${\hat
g}$.
The effect, or back-reaction, of the quantum fields on the
geometry described by the metric
${\hat g}$ induces a modification of this metric, $\delta g$, so that

${\hat g} \rightarrow {\hat g} + \delta g$ when we ``turn on'' the
sources.
These quantum stress tensors obey the background conservation law
\begin{equation}
{\hat \nabla}_{\mu}\, <T^{\mu \nu}> = 0,
\end{equation}
with respect to the background covariant derivative:
${\hat \nabla} \sim \partial + {\hat \Gamma} ({\hat g})$.
As a consequence, it turns out we may only solve for the modified
metric to order $O(\hbar)$. This we can demonstrate easily by
expanding the Einstein tensor $G$
in powers of the Planck constant and
 using (1) and (3), thus leading to
\begin{equation}
{\hat \nabla}_{\mu}\, ({\hat G} + \delta G + \Delta G )^{\mu \nu} =
 8\pi \, {\hat \nabla}_{\mu} \, <T^{\mu \nu}> = 0,
\end{equation}
so that ${\hat \nabla}_{\mu}(\delta G + \Delta G)^{\mu \nu} = 0$, or
to order $\hbar$, ${\hat \nabla}_{\mu} \, \delta G^{\mu \nu} = 0$.
Here, $\delta G$ and $\Delta G$ denote the order $\hbar$ and
higher-order contributions, respectively.
Since the operator ${\hat \nabla}$ is $O(1)$ while the stress tensor
is $O(\hbar)$, the back-reaction
equation (2) reduces to
\begin{equation}
\delta G^{\mu \nu}({\hat g} + \delta g) = 8\pi \, <T^{\mu \nu}>.
\end{equation}
The components of the stress tensors are treated as input. To solve
(5), one inserts the most general metric ansatz compatible with the
symmetries and coordinate dependence of $<T_{\mu \nu}>$ into the
left hand side.

Turning now to the case at hand, we consider the black hole background,
whose metric is
\begin{equation}
{\hat g}_{\mu \nu} = {\rm diag}\,\left( -(1-\frac{2M}{r}),
(1-\frac{2M}{r})^{-1}, r^2, r^2 \sin^2 \theta \right),
\end{equation}
and $M$ is the mass of the black hole.
As shown in \cite{York85,HKYa}, the modified
metric, which is static and spherically
 symmetric, can be put into the following form:
\begin{equation}
ds^2 = -\left( 1 - \frac{2m(r)}{r} \right) (1 + 2 \epsilon {\bar
\rho}(r)) \, dt^2 + \left( 1 - \frac{2m(r)}{r} \right)^{-1} dr^2 + r^2
d\Omega^2,
\end{equation}
where $m(r), {\bar \rho}(r)$ are two functions depending on $r$ and
$\epsilon = (M_{Planck}/M_{BH})^2 < 1$ is the expansion parameter.
Note that $\epsilon = \hbar/M^2 $ in units where $G=c=k_B=1$. This
parametrization of the new equilibrium metric reflects the fact that
the back reaction induces static, spherically symmetric metric
perturbations. Indeed, the stress tensors renormalized over
$\hat g$ are static and depend only on the radial coordinate.
$d\Omega^2$ is the standard metric of a normal round unit sphere.
The mass function has the form $m(r) = M[1 + \epsilon(\mu(r) +
CK^{-1})]$ where $C$ is a constant of integration which serves to
renormalize the bare Schwarzschild mass $M$. Indeed, to the order we
are working, we may write
\begin{equation}
m(r) = M \left( 1 + \epsilon C K^{-1} \right)[1 + \epsilon \mu(r)]
\equiv M_{ren}[1 + \epsilon \mu(r)].
\end{equation}
We henceforth write $M$ for the black hole mass in what
follows, with the understanding that this
represents the physical black hole mass. We note from (11) that
$M_{rad} = \epsilon M \mu(r)$ is the usual expression for the
effective mass of a spherical source. The metric in (7) is completed by the
determination of $\bar \rho$, where
\begin{equation}
{\bar \rho}(r) = \rho(r) + kK^{-1},
\end{equation}
with $k$ another constant of integration; $K=3840\pi$.

The corresponding semiclassical field equations, valid to
$O(\epsilon)$
are $(w = 2M/r)$
\begin{eqnarray}
\epsilon \frac{d {\rho}}{dw} & = & -\frac{16\pi M^2}{w^3} (1 -
w)^{-1} <T^r_r - T^t_t> \nonumber \\
\epsilon \frac{d \mu}{dw} & = & \frac{32\pi M^2}{w^4} <T^t_t>.
\end{eqnarray}
These may be obtained substituting (7) into (5) keeping only the
$O(\epsilon)$ terms. Naturally, indices are raised/lowered with the
background metric.

Once the indicated components of the renormalized stress-energy
tensors are known, the solution of the semiclassical
back-reaction equation (5)
follows immediately from two simple integrations:
\begin{eqnarray}
\mu(r) &=& \frac{1}{\epsilon M} \, \int_{2M}^r <-T^t_t>\, 4\pi {\tilde
r}^2\, d{\tilde r}, \\
\rho(r) &=& \frac{1}{\epsilon} \, \int_{2M}^r <T^r_r - T^t_t>\,
4\pi {\tilde r}^2 \, d{\tilde r}.
\end{eqnarray}
Actually, the back reaction problem as it
stands has no \underline{definite}
solution
 unless boundary conditions are specified \cite{York85}.
There are a number of reasons for why this must be so. In the first
instance, the constant of integration $k$, appearing in ${\bar \rho}$
remains undetermined unless a boundary condition is invoked
(asymptotic flatness does $\it not$
 fix this constant \cite{York85}). More
importantly, the renormalized stress tensors employed are
asymptotically {\em constant}, thus the radiation in a sufficiently
large spatial region would collapse onto the black hole thereby
producing a larger one. It is therefore necessary to implant the
system consisting of the black hole plus radiation in a finite cavity
with wall radius $= r_o > 2M$. As discussed in \cite{York86}, a very important
consequence
of considering black holes in spatially bounded regions, quite
independent of the back reaction problem, is that the cavity stabilizes
the black hole in the thermodynamic sense
and yields a {\em positive} heat capacity for the hole.
There are at least two distinct types of
physically relevant boundary conditions one may
choose to impose. In the case of canonical boundary conditions,
we specify the temperature of the cavity wall $T(r_o)$ and immerse the
cavity containing the black hole and radiation in an
external heat reservoir whose temperature $T = T(r_o)$.
To obtain microcanonical boundary conditions, we specify the total
 energy at the cavity wall $E(r_o)$ and match on an external
Schwarzschild spacetime with effective mass $m(r_o)$.  In the former
case, the integration constant is absorbed by a redefinition of the
time coordinate. This is possible as coordinate time has no special
meaning unless the metric is asymptotically constant. We can choose
the timelike Killing vector to be $\frac{\partial}{\partial {\bar t}}
= \lambda \frac{\partial}{\partial t}$ for $\lambda$ a constant. The
choice $\lambda = (1 -\epsilon k K^{-1})$ removes $k$ from expressions
for the physical quantities.
In the latter case, we fix $k$ by requiring continuity
of the metric across the cavity wall: $k = -\rho(r_o) K$.

How does one determine the size of the cavity? The answer to this
question is tied
up with the perturbative validity of the solutions, which may be
maintained provided the cavity wall radius satisfies a certain
inequality, to which we now turn. We recall that all the stress
tensors renormalized on the black hole background satisfy
\cite{Howard84,Jensen89,Brown86}
\begin{equation}
<T^{\mu}_{\nu}> {\longrightarrow}
({\rm spin-dependent \,\, const.}) \times {\rm diag} (-3,1,1,1)^{\mu}_{\nu},
\end{equation}
as $r \rightarrow \infty$,
which results in asymptotically unbounded metric perturbations
$\delta g_{\mu \nu} = (g_{\mu \nu} - {\hat g}_{\mu \nu})$.
In fact, one can show \cite{York85,HK93,HKYa, HKYb} that
the relative corrections diverge
quadratically
\begin{equation}
|\frac{\delta g}{\hat g}| {\longrightarrow} = \epsilon \, \alpha_s
\left( \frac{2}{3K} \right) \left( \frac{r}{2M} \right)^2,
\end{equation}
($tt$ and $rr$ components only, since $\delta g_{\theta \theta}=
\delta g_{\phi \phi} = 0$)
with $\alpha_s$
 a spin-dependent constant: $\alpha_s = (1/2, 7/8, 1)$ for the spins
$s = (0,1/2,1)$, respectively.
So, we can obtain solutions that are uniformly small over the entire
range $2M < r < r_{domain}$, taking $|\frac{\delta g}{\hat g}| < \delta
< 1$, which from (14), and by saturating the previous inequality,
defines the radius of the domain of perturbative
validity via
\begin{equation}
\left( \frac{r_{domain}}{2M} \right)^2 = \frac{3K}{2 \alpha_s}
\left( \frac{\delta}{\epsilon} \right).
\end{equation}
So, we should take the cavity radius $r_o < r_{domain}$. By way of
illustration, setting $\epsilon = \delta < 1$ results in rather large
perturbatively valid domains, indeed $r_{domain} \approx (380M, 286M,
268M)$ for the spins $s = (0,1/2,1)$, respectively.

\section{The Quantum Stress-Energy Tensors}
These have been obtained in exact form for the conformal scalar \cite{Howard84}
and U(1) gauge boson \cite{Jensen89}, and in an approximate
form for the massless
spin 1/2 fermion \cite{Brown86}. For the former
two cases, they are expressed as
\begin{equation}
<T^{\mu}_{\nu}>_{renormalized} = <T^{\mu}_{\nu}>_{analytic}
 + \left( \frac{\hbar}{\pi^2 (4M)^4} \right) \, \Delta^{\mu}_{\nu},
\end{equation}
where the analytic piece, in the case of the conformal scalar, was
first given by Page \cite{Page82}. The term
$\Delta^{\mu}_{\nu}$ is obtained from
a numerical mode sum. As this term is small in comparison to the
analytic piece, we do not include it here. This affects none of our
qualitative results; both pieces seperately obey the the required
regularity and consistency conditions. The analytic part has the exact
trace anomaly in both cases. We display only the analytic part for the
U(1) case and direct
the interested reader to the original literature for the other cases.
Dropping the angular brackets, we have ($w = 2M/r$)
\begin{eqnarray}
T^t_t & = & -\frac{1}{3}aT^4_H \left(3+6w+9w^2+12w^3-315w^4+78w^5-249w^6
\right) , \nonumber  \\
T^r_r & = & \frac{1}{3}aT^4_H
\left(1+2w+3w^2-76w^3+295w^4-54w^5+285w^6 \right), \nonumber \\
T^{\theta}_{\theta} = T^{\phi}_{\phi} & = & \frac{1}{3}aT^4_H
\left(1+2w+3w^2+44w^3-305w^4+66w^5-579w^6  \right), \nonumber
\end{eqnarray}
where $a = (\pi^2/15{\hbar}^3)$ and $T_H = {\hbar}/{8\pi M}$ is the
(uncorrected) Hawking temperature.

\section{Solutions}

The explicit solutions to the $O(\epsilon)$ back reaction (5) obtained
using the
above stress-tensors may be summarized compactly as follows.  Denoting
with the subscripts $S,f,V$ the conformal scalar, massless fermion and
vector boson respectively, the metric functions in (11,12) turn out to be
\cite{York85,HK93,HKYa,HKYb}
\begin{eqnarray}
K \mu_S & = & \frac{1}{2} \left[ \frac{2}{3} w^{-3} +2w^{-2} +
 6w^{-1}- 8\ln(w)-10w-6w^2+22w^3 -\frac{44}{3} \right], \nonumber \\
 K \rho_S & = & \frac{1}{2} \left[ \frac{2}{3} w^{-2} +4w^{-1} -8\ln(w)
 -\frac{40}{3}w -10w^2 -\frac{28}{3}w^3 +\frac{84}{3} \right],
\end{eqnarray}
for the conformal scalar,
\begin{eqnarray}
K \mu_f & = & \frac{7}{8} \left[ \frac{2}{3}w^{-3} +2w^{-2}+6w^{-1}-8\ln(w)
-\frac{90}{7}w-\frac{62}{7}w^2+\frac{46}{3}w^3-\frac{16}{7} \right],
\nonumber \\
K \rho_f & = & \frac{7}{8}
\left[ \frac{2}{3}w^{-2}+4w^{-1}-8\ln(w)-\frac{200}{21}w
-\frac{50}{7}w^2-\frac{52}{7}w^3+\frac{136}{7} \right],
\end{eqnarray}
for the massless spinor, and
\begin{eqnarray}
K \mu_V & = & \frac{2}{3}w^{-3} +2w^{-2} +6w^{-1} -8\ln(w) +210w
-26w^2 + \frac{166}{3}w^3 -248, \nonumber \\
K \rho_V & = & \frac{2}{3}w^{-2} +4w^{-1} -8\ln(w) + \frac{40}{3}w
+ 10w^2 + 4w^3 -32,
\end{eqnarray}
for the U(1) gauge boson. The various spin cases are distinguished by
the spin dependence of the numerical coefficients.

\section{Corrected Black Hole Temperature}

As is well known, a black hole in empty space radiates quanta
possessing a temperature characterized by the mass $M$ of the hole.
At large distances from the hole $(r >> M)$, the temperature of the
radiation approaches $T_{\infty} = {\hbar}/{8\pi M} \equiv T_H$.
The presence of this
radiation however, leads to modifications of this temperature when the
back reaction is properly taken into account. For microcanonical
boundary conditions, the corrected temperature at spatial infinity
takes the form
\begin{equation}
T_{\infty} = \frac{\hbar}{8\pi M} \left(1 + \epsilon\, f(r_o,s,M)
\right),
\end{equation}
where $f$ is a calculable function of the cavity radius $r_o$, the
spin $s$ of the
quantum radiation and the renormalized mass $M$ of the hole.
Note that generally,
$T_{\infty}$
is {\em not} equal to the Hawking temperature.

To find the form of $f$ we recall that
\begin{equation}
T_{\infty} = \frac{\hbar \,\kappa_H}{2\pi},
\end{equation}
where $\kappa_H$ is the surface gravity of the event horizon; for a
``free'' black hole, $\kappa_H = (4M)^{-1}$. In the case of back
reaction, the surface gravity is given by \cite{York85}
\begin{equation}
\kappa_H = \frac{1}{4M} \left( 1 + \epsilon ({\bar \rho} - \mu)
 + 8\pi r^2 \, <T^t_t> \right)_{r=2M}.
\end{equation}
With the microcanonical boundary
conditions, one then obtains \cite{York85,HKYa}
\begin{equation}
T^{(s)}_{\infty} = \frac{\hbar}{8\pi M} \left(1 - \epsilon \rho_s(r_o)
+ \epsilon n_s K^{-1} \right),
\end{equation}
where the constant $n_s = (12, -4, 304)$ for the spin $s = (0,1/2,1)$.
The local temperature, $T_{loc}$, is obtained by
blueshifting back from infinity
to a finite value of $r$
(by means of the Tolman factor \cite{Tolman34}):
\begin{equation}
T_{loc}(r) = T_{\infty} [-g_{tt}(r)]^{-1/2}.
\end{equation}
This yields the local temperature valid for all $r > 2M$
(dropping the spin labels)
\begin{equation}
T_{loc}(r) = T_H (1 - w)^{-1/2} \left[ 1 + \epsilon( \rho(r)-nK^{-1}
-\frac{w}{2} (1 - w)^{-1}\, \mu(w) ) \right].
\end{equation}

\section{Thermodynamical Entropy}

One way to calculate the entropy is as follows. Consider first the
case of the free black hole. From the 1st law of thermodynamics
applied to slightly differing equilibrium systems
\begin{equation}
dE = dQ + dW,
\end{equation}
and as no work is done on the system, $dW = 0$, so that
\begin{equation}
dS = \frac{dQ}{T} = \frac{dE}{T} = \frac{dM}{T_{\infty}},
\end{equation}
where $M$ is the black hole mass and $T_{\infty}$ the temperature at
spatial infinity. Integrating, we have that the entropy
\begin{equation}
S = \int \frac{dM}{T_{\infty}} = \int (\frac{8\pi M}{\hbar}) dM =
\frac{4\pi M^2}{\hbar} + {\rm const.}
\end{equation}
This yields the usual Bekenstein-Hawking expression for the black
hole entropy after setting the constant of integration to zero:
$S=S_{BH}$.

 Now turn on the back reaction and compute $S$ from (28). We must of
course use the corrected asymptotic temperature (23) and the
effective mass of the combined system of black hole plus radiation:
$m(r_o) = M[1 + \epsilon \mu(r_o)]$ in (28).
With (remembering that $\epsilon = \epsilon(M)$)
\begin{equation}
dm(r_o) = \left[1 + \epsilon\,(w \frac{\partial \mu}{\partial w} -
\mu(w)) \right]_{r_o}\, dM ,
\end{equation}
we obtain $(dr_o = 0)$ that
\begin{equation}
S = S_{BH} + \Delta S(r_o) + {\rm constant}.
\end{equation}
The constant of integration is fixed by
requiring that $\Delta S(r_o = 2M) = 0$, that
is, with no ``room'' for the fields to contribute anything further,
one should obtain just the Bekenstein-Hawking entropy
$\frac{1}{4}A_H/{\hbar}$, as would be expected. With this choice, we
find \cite{HKYb} for the conformal scalar field
\begin{eqnarray}
 \Delta S_S = \frac{8\pi}{K}(\frac{1}{2}) ( \frac{8}{9}w^{-3} +
 \frac{8}{3}w^{-2} + 8w^{-1} + \frac{32}{3}\ln (w) & - & \frac{40}{3}w -
 8w^2  \nonumber \\
 & + &  \frac{104}{9}w^3-\frac{16}{9} ),
\end{eqnarray}
while for the massless spin-$\frac{1}{2}$ fermion
\begin{eqnarray}
\Delta S_f = \frac{8\pi}{K}(\frac{7}{8}) ( \frac{8}{9}w^{-3}
+\frac{8}{3}w^{-2} +8w^{-1} + \frac{128}{7} \ln (w) & - & \frac{200}{21}w
-8w^2 \nonumber \\
 & + & \frac{488}{63}w^3 -\frac{16}{9} ),
\end{eqnarray}
and
\begin{eqnarray}
\Delta S_V = \frac{8\pi}{K} ( \frac{8}{9}w^{-3} +
\frac{8}{3}w^{-2} +8w^{-1} -96 \ln (w) & + & \frac{40}{3}w-8w^2
\nonumber \\
 & + & \frac{344}{9}w^3-\frac{496}{9} ),
\end{eqnarray}
for the U(1) gauge field.
Among the features enjoyed by these entropy corrections, it is
important to note that they all are positive definite, $\Delta S \geq
0$, and are monotone increasing functions of $r_o$ . They are thus
amenable to arguments relating thermodynamical to statistical entropy.
Moreover, the corrections are of order $O(1)$ in $\hbar$ and so are of
the same order as the naive flat space radiation entropy:
$S_{flat} = \frac{4}{3}aT^3_H V$. These corrections are therefore important.

\section{Concluding Remarks}

We should like to comment on the following points.

Lowest order solutions of the semi-classical back reaction problem
have been used to calculate the $O(\hbar)$ corrections to the
temperature and the order $O(1)$ correction to the
entropy of a Schwarzschild black hole.
It should be emphasized that this is a perturbative calculation. It
 would be of interest to extend the results to higher order, although
a non-perturbative treatment would clearly be preferable.
At higher orders, one needs to be careful, however, because the
semi-classical field equation may well receive corrections arising
from the stress tensor renormalization:
\begin{equation}
G_{\mu \nu} + \Lambda g_{\mu \nu} + a H^{(1)}_{\mu \nu}
 + b H^{(2)}_{\mu \nu} = 8\pi G <T_{\mu \nu}>,
\end{equation}
where $\Lambda, a, b$ are constants and the tensors
$H^{(1)}_{\mu \nu}$ and $H^{(2)}_{\mu \nu}$ are linear combinations
of quadratic curvature terms \cite{Birrell82, Grib94}.

Perhaps of greater importance is the fact that the contribution of
quantum metric fluctuations has not been taken into account.
Provided the semi-classical back reaction program leads qualitatively
in the right direction (that is, towards a correct quantum gravity),
one should include the spin-2 graviton contribution to the effective
stress-energy tensor. One in fact expects the effects of linear
gravitons to contribute a term of the same order to the stress tensor
as those coming from ordinary matter and radiation fields.
The corresponding calculations could be carried out
once a renormalized effective
energy-momentum tensor for quantized linear metric perturbations
over a black hole background is
worked out. However, at present, certain technical obstructions appear to make
this a difficult task \cite{Jensen95}.

Lastly, one can use the results of the back reaction problem to
explore the nature of the modified black hole spacetime by means of
the effective potential \cite{HKYb}, which determines the motion of
test particles in the vicinity of the perturbed black hole,
and to estimate the black
hole free energy, of particular importance for assessing under
what conditions the nucleation phase transition of black holes
from hot flat space is likely to occur \cite{Hochberg95}.

\vspace{1cm}
\noindent
{\bf Acknowledgements}
This work is based on collaborations with Thomas Kephart and James W.
York, Jr. I thank Dr. Michael Bordag
of Leipzig University for providing me with the partial
financial support which facilitated my attending this Workshop.

\end{document}